\documentclass[aps,prd,reprint,superscriptaddress,showpacs]{revtex4-2}
\usepackage{amsmath,graphicx,subfigure,amssymb}
\usepackage{changes} 
\usepackage[toc,page]{appendix}
\usepackage{multirow,longtable}

\newcommand*\diff{\mathop{}\!\mathrm{d}}
\newcommand*\Diff[1]{\mathop{}\!\mathrm{d^#1}}

\begin{document}


\title{Dependence of Electrostatic Patch Force Evaluation on the Lateral Resolution of Kelvin Probe Force Microscopy}

\author{Kun Shi}
\affiliation{School of Physics, Huazhong University of Science and Technology, Wuhan 430074, China}

\author{Pengshun Luo}
\email[E-mail: ]{pluo2009@hust.edu.cn}
\affiliation{MOE Key Laboratory of Fundamental Physical Quantities Measurement, Hubei Key Laboratory of Gravitation and Quantum Physics, PGMF and School of Physics, Huazhong University of Science and Technology, Wuhan 430074, China}

\author{Jinquan Liu}
\affiliation{MOE Key Laboratory of Fundamental Physical Quantities Measurement, Hubei Key Laboratory of Gravitation and Quantum Physics, PGMF and School of Physics, Huazhong University of Science and Technology, Wuhan 430074, China}

\author{Hang Yin}
\affiliation{MOE Key Laboratory of Fundamental Physical Quantities Measurement, Hubei Key Laboratory of Gravitation and Quantum Physics, PGMF and School of Physics, Huazhong University of Science and Technology, Wuhan 430074, China}

\author{Zebing Zhou}
\affiliation{MOE Key Laboratory of Fundamental Physical Quantities Measurement, Hubei Key Laboratory of Gravitation and Quantum Physics, PGMF and School of Physics, Huazhong University of Science and Technology, Wuhan 430074, China}


\date{\today}

\begin{abstract}
Kelvin Probe Force Microscopy (KPFM) is widely used to measure the surface potential on samples, from which electrostatic patch force can be calculated. However, since the KPFM measurements represent a weighted average of local potentials on the sample, the accuracy of the evaluation critically depends on the precision and lateral resolution of the method. In this paper, we investigate the influence of this averaging effect on patch force estimations using both analytic and numerical methods. First, we derive the correlation functions of patch potential and establish the formulas for calculating the electrostatic patch forces in the parallel-plate geometry, with and without consideration of the KPFM measurement effect. Thus, an analytic method is established to determine the accuracy of patch force evaluation when the statistical parameters of the patch potential and the lateral resolution of the KPFM are given. Second, numerical simulations are employed to explore the dependence of estimated patch forces on the KPFM's lateral resolution under more realistic conditions. Both analytic and numerical results show a similar dependence of the patch force estimation on the patch characteristic size, potential fluctuation and the lateral resolution of the KPFM. It is also found that the underestimation of the patch force becomes less sensitive to the KPFM's resolution as the separation between plates increases. The results of this study could provide useful guidance for the accurate evaluation of electrostatic patch forces using KPFM.
\end{abstract}


\maketitle

\section{Introduction}

Potential fluctuations on sample surfaces, known as the patch effect, are a common source of error in experiments involving force measurements between closely spaced objects \cite{camp_macroscopic_1991}. Initially predicted by the theory of anisotropy in the electronic work function of metals \cite{smoluchowski_anisotropy_1941}, the patch effect has since been observed in various experimental settings \cite{rossi_observations_1992,vatel_kelvin_1995,gaillard_method_2006,ferroni_electrostatic_2011,yin2014measurements,li2022precision,li2023analysis,song2023high}. The primary origin of the patch effect is the variation in crystallographic directions on metal surfaces \cite{speake_forces_2003}. Additional mechanisms contributing to patch effects include variations in chemical composition, nonuniform dipole layers on the surface, and surface contaminations\cite{ferroni_electrostatic_2011,ke_electrostatic_2023}.

The patch effect can induce random electrostatic field noise, serving as sources of errors in various high-precision experiments, such as ion traps \cite{wineland_experimental_1998,hite_surface_2021}, atomic force sensing measurements \cite{stipe_noncontact_2001}, and Casimir force measurement experiments \cite{garrett_measuring_2020,garrett_effect_2015,lamoreaux_demonstration_1997,mohideen_precision_1998,kim_anomalies_2008,bimonte_isoelectronic_2016,decca_precise_2005,woods_materials_2016,sukenik_measurement_1993,behunin2014kelvin,behunin2014limits,behunin2012electrostatic}. A particularly significant example where the patch effect poses a major concern is in space-based gravitational wave detection  \cite{robertson_kelvin_2006,lockerbie_gravitational_1996,armano_charge-induced_2017,schumaker_disturbance_2003,weber_position_2003,sun_grating_2006,Luo_2016}. These space missions typically employ multiple free-fall test masses housed in drag-free spacecrafts to ensure that the motion of the test masses is solely influenced by gravitational interactions. Nevertheless, spatial variations in the surface potential of the test masses and their enclosures introduce electrostatic force noise, which can lead to measurement errors\cite{speake_forces_1996}.

The magnitude of the patch force can be determined from the surface potential \cite{speake_forces_1996}. Therefore, accurately measuring the surface potential with KPFM at an appropriate resolution is crucial in quantifying the patch effect \cite{speake_forces_2003}. In gravitational wave detection missions, such as LISA Pathfinder, the separation between the 46-mm cubes of test mass and its housing ranges from 2.9 mm to 4 mm \cite{armano_charge-induced_2017}. The large surface area of the test mass ($\sim2000\,\text{mm}^2$) requires careful consideration of the lateral resolution, as higher resolution would significantly increase scan time without proportionally improving the patch force estimation accuracy. In the KPFM measurements reported in Ref. \cite{robertson_kelvin_2006}, an actual lateral resolution of $\sim$ 1 mm was used, and it took one hour to scan a $100\,\text{mm}^2$ area. Since this resolution is smaller than the sample size and the separation, it was considered sufficient to estimate the patch force. In short-range experiments, patch effects remain a significant source of systematic uncertainty \cite{speake_forces_2003, behunin2012electrostatic}. For example, in Casimir force measurements, two interacting surfaces are usually several hundred nanometers apart\cite{behunin2014kelvin}. The surface potential of the sample needs to be measured independently to accurately subtract the electrostatic force contribution in such measurements\cite{behunin2012electrostatic}. To achieve this, KPFM with a lateral resolution below 100 nm is commonly used. However, the optimal choice of the resolution has not been thoroughly studied, and further investigations are needed to assess its influence on accurately evaluating the patch force.

In this paper, we investigate the effect of the lateral resolution of KPFM on the patch force evaluation and provide guidance for accurately assessing patch force in related experiments. We derive the correlation functions of the potential and obtain the analytic integrals to calculate the patch forces, with and without consideration of the KPFM measurement effect. Numerical simulations under more realistic situations are also carried out to compare with the analytic methods, and similar results are obtained.

The structure of this paper is as follows:
In Section II, we derive the autocorrelation functions of the patch potential, inspired by the quasi-local correlation model and Buffon-Laplace needle problem, and develop the formulas to calculate the patch forces. In Section III, we utilize finite element simulations to calculate the electrostatic force and compare the results with the results obtained by the analytic method, investigating the dependence of patch force evaluation on the KPFM's resolution under different conditions. In Section IV, we summarize our findings and discuss potential extensions of this work.

\section{Theoretical Framework}

\begin{figure}
	\includegraphics{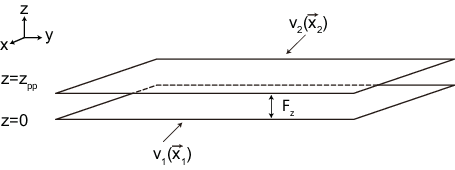}
	\caption{
		Electrostatic Force $F_z$ between the plates  is determined by the boundary potential $v_1(\vec{x}_1)$ and $v_2(\vec{x}_2)$.
	}
	\label{fig:Plate_Plate_Model}
\end{figure}

\subsection{Calculation of the Electrostatic Patch Force for Given Potential}

Before proceeding with our analysis, we briefly review the established methodologies for calculating the electrostatic patch forces $F_z$ between two infinite, parallel plates separated by a distance $z_{pp}$ (Fig.\ \ref{fig:Plate_Plate_Model}). Following this, we derive our own correlation functions to obtain the analytic expressions for the patch force.

In the infinite plate-plate geometry, each surface has an independent potential distribution, $v_1(\vec{x})$ and $v_2(\vec{x})$, both with a mean of zero and a root mean square (RMS) value of $V_{rms}$. The potential correlation function is defined as
\begin{equation}
	c_{i,j}(\vec{x})=\left<v_i (\vec{\xi}) v_j (\vec{\xi} - \vec{x}) \right>_v,
\end{equation}
where $i$ and $j$ are the plate numbers.

\begin{figure}
	\includegraphics{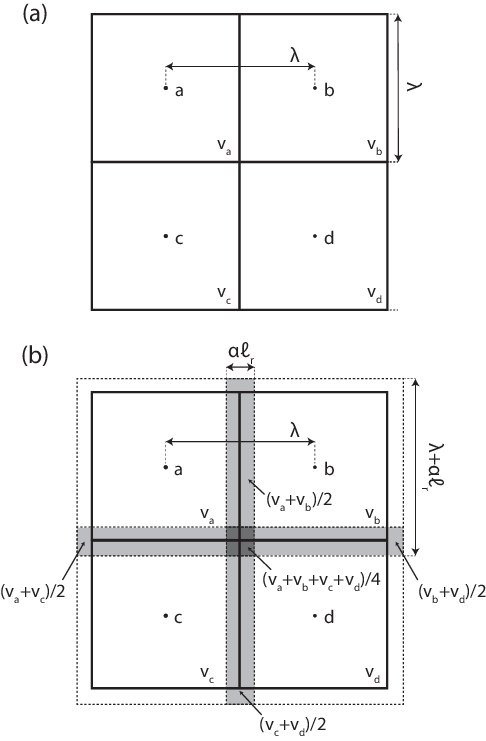}
	\caption{
		Illustration of the patch models for the derivation of correlation functions. (a) Square patches with a side length of $\lambda$. (b) Boundary zones (grey areas) are introduced in the square patch model. The potential in the boundary zones is the average of the adjacent patches.
	}
	\label{fig:Patch_Model}
\end{figure}

Following the theoretical framework established by Speake and Trenkel \cite{speake_forces_1996}, the electrostatic patch force per unit area between the plates is derived as
\begin{equation}
	F_z=-\frac{\epsilon_0}{(2\pi)^2} \iint \frac{k^2 C(\vec{k})}{\sinh^2 k z_{pp}} \Diff2 \vec{k}.
	\label{eq:Patch_Force_Equation}
\end{equation}
Note that the potential distributions on different plates are independent, thus $C_{12}(\vec{k})=0$. Since the potentials on different plates obey the same distribution, we can define $C(\vec{k}) = C_{11}(\vec{k}) = C_{22}(\vec{k})$.

The calculation of the electrostatic patch force depends on the model used to describe the potential distribution. Here, we adopt the quasi-local correlation model to perform the following analysis, as it closely approximates the actual distribution of patch potentials \cite{behunin_modeling_2012}. In this model, each patch has a potential that follows an identical and independent distribution. The surface potential can be expressed as \cite{behunin_modeling_2012}
\begin{equation}
	V(\vec{x})=\sum_{a} v_a \Theta_a (\vec{x}),
	\label{eq:Potential_Assignment}
\end{equation}
where $v_a$ represents a random variable corresponding to the potential of the \textit{a}-th patch, and  $\Theta_a(\vec{x})$ is an indicator function that equals to 1 for points $\vec{x}$ on the \textit{a}-th patch and 0 for points outside the patch. Using $\left< v_a v_b \right>_v = \delta_{ab} V_{rms}^2$, the correlation function for the potentials at two points is derived as
\begin{equation}
	C(\vec{x},\vec{x}^{'}) = V_{rms}^2 \sum_a \Theta_a (\vec{x}) \Theta_a (\vec{x}^{'}),
\end{equation}
where $\Theta_a (\vec{x}) \Theta_a (\vec{x}^{'})$ is nonzero only when both $\vec{x}$ and $\vec{x}^{'}$ are located within the same \textit{a}-th patch. 

We further introduce the patch pair coverage function $E(\vec{x},\vec{x}^{'})$, defined as
\begin{equation}
	E(\vec{x},\vec{x}^{'}) = \sum_a \Theta_a (\vec{x}) \Theta_a (\vec{x}^{'}),
\end{equation}
which represents the number of patches containing both points $\vec{x}$ and $\vec{x}^{'}$. Assuming translation and rotation invariance, the correlation function depends only on the distance between the two points. As a result, the patch pair coverage function is simplified to $E(r)$, which is the expected number of patches containing any two points at a distance of $r$ apart. The correlation function is then given by
\begin{equation}
	C(r) = V_{rms}^2 \cdot E(r).
	\label{eq:Correlation_Function}
\end{equation}

To derive an analytical expression of $E(r)$, we simplify the model by assuming that the patches are squares with a side length of $\lambda$  [Fig.\ \ref{fig:Patch_Model}(a)]. This allows us to utilize the results from the Buffon-Laplace Needle Problem, where the probability of a randomly positioned needle not intersecting any line in a grid is calculated \cite{laplace1820theorie}. Since the condition ``needle does not intersect any line'' is equivalent to the condition ``one patch contains both end points'', the expression for $E(r)$ is obtained as
\begin{equation}
	E (r)=
	\begin{cases} 
	 	1 - \frac{\frac{r}{\lambda} \times \left( 4 - \frac{r}{\lambda} \right)}{\pi} & \text{for } 0 \leq r \leq \lambda \\
		0 & \text{for } r > \lambda
	\end{cases}.
	\label{eq:Without_KPFM}
\end{equation}
Thus, the correlation function is given by
\begin{equation}
	C(r) = V_{\text{rms}}^2 \left( 1 - \frac{\frac{r}{\lambda} \times \left( 4 - \frac{r}{\lambda} \right)}{\pi} \right).
	\label{eq:Correlation_Without_KPFM}
\end{equation}

With the above correlation function and Eq.\ \eqref{eq:Patch_Force_Equation}, we derive the formula for the electrostatic patch force between two infinite parallel surfaces characterized by $z_{pp}$, $\lambda$ and $V_{rms}$, which can be expressed as
\begin{widetext}
	\begin{equation}
		F_z = - \frac{\epsilon_0 V_{rms}^2}{z_{pp}^2} \int_0^{\infty} \frac{k}{\sinh^2 k} \left( \int_0^{k \frac{\lambda}{z_{pp}}} r J_0(r) \left( 1-\frac{\frac{r}{k \frac{\lambda}{z_{pp}}}\times (4 - \frac{r}{k \frac{\lambda}{z_{pp}}})}{\pi}\right) \diff r \right) \diff k,
		\label{eq:Force_z}
	\end{equation}
\end{widetext}
where $J_0$ denotes the 0th-order Bessel function. It is important to note that the integral depends solely on the parameter $ z_{pp}/\lambda$. When $ z_{pp} \ll \lambda$, it can be approximated as 1, thus $F_z$ approaches the limit of $\epsilon_0 V_{rms}^2 / z_{pp}^2$.
This result aligns with Eq.\,(6) in Ref.\,\cite{behunin_modeling_2012}, as in this scenario, the patch potential in the plate-plate model effectively reduces to that of a parallel plate capacitor with uniform potential, following the distribution of $V_{rms}$ on both surfaces.

\subsection{Calculation of Electrostatic Patch Force For KPFM-Measured Potential}

To account for the averaging effect in KPFM measurements, we assume a boundary zone between adjacent patches where the potential is averaged, as shown by the shaded area in Fig.\ \ref{fig:Patch_Model}(b). The width of this boundary ($\alpha \ell_r$) is set to be proportional to the KPFM resolution ($\ell_r$). The inclusion of the coefficient $\alpha$ is necessary since the definition of KPFM resolution does not directly account for this boundary zone width (discussed in detail in Appendix \ref{appendix:linearity}). This assumption holds when $\ell_r < \lambda$. Thus, the KPFM-measured potential can be described as
\begin{equation}
	V(\vec{x})=\frac{\sum_a v_a \Theta_a(\vec{x})}{\sum_b \Theta_b(\vec{x})},
	\label{eq:Surface_Potentail_KPFM}
\end{equation}
where the patches summed are those with a side length of $\lambda + \alpha \ell_r$, as shown by the dotted squares in Fig.\ \ref{fig:Patch_Model}(b). The denominator represents the total number of overlapping patches, while the numerator sums the potentials of these patches. This expression gives the averaged potential within the boundary zone.

The correlation function for the potentials at two points is then given by
\begin{equation}
	C(\vec{x},\vec{x}^{'})=V_{rms}^2 \cdot \frac{\sum_a \Theta_a(\vec{x}) \Theta_a(\vec{x}^{'})}{ \left( \sum_b \Theta_b(\vec{x}) \right) \cdot \left( \sum_c \Theta_c(\vec{x}^{'}) \right) }.
	\label{eq:Correlation_KPFM}
\end{equation}
Assuming translation and rotation invariance, we obtain the potential correlation function using the patch pair coverage function introduced in Sec. II A as
\begin{equation}
	C(r)=V_{rms}^2 \cdot \frac{E(r)}{E^2(0)}.
	\label{eq:Correlation_Simplified_KFPM}
\end{equation}
The detailed derivation is provided in Appendix \ref{appendix:Derivation}. 

\begin{figure}
	\includegraphics{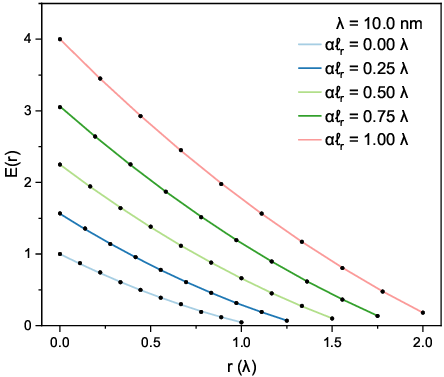}
	\caption{Comparison of the patch pair coverage function obtained by Monte-Carlo simulation (solid dots) and Eq.\ \eqref{eq:Correlation_Function_2} (solid lines) for different values of $\alpha \ell_r$.}
	\label{fig:Emperical_Function}
\end{figure}

The patch pair coverage function $E(r)$ is revised based on the following arguments: (1) As shown in Fig.\  \ref{fig:Patch_Model}(b), the side length of each patch is changed from $\lambda$ to $\lambda + \alpha \ell_r$ while the center of the patch remains fixed. (2) The area of each patch is effectively expanded by a factor of $(1+\frac{\alpha \ell_r}{\lambda})^2$, leading to patch overlapping. This overlap between patches causes the patch pair coverage function to exceed 1, increasing approximately in proportion to the rate of area increment. Considering these factors, the patch pair coverage function is then revised as
\begin{equation}
	E (r) = \left( 1 + \frac{\alpha \ell_r}{\lambda} \right)^2 \times \left( 1 - \frac{\frac{r}{\lambda+\alpha \ell_r} \times \left( 4 - \frac{r}{\lambda + \alpha \ell_r} \right)}{\pi} \right).
	\label{eq:Correlation_Function_2}
\end{equation}

We employ Monte Carlo simulations to verify this formula. To calculate the value of $E(r)$ for a specific $r$, needles of length $r$ with varying orientations and positions are randomly generated on the patch model depicted in Fig.\ \ref{fig:Patch_Model}(b). For each needle, the number of patches containing the full needle is counted, and the average value is calculated to determine $E(r)$. By repeating this procedure for different values of $r$, we obtain the patch pair coverage function $E(r)$. Figure \ref{fig:Emperical_Function} shows that the simulation results are in good agreement with Eq.\ \eqref{eq:Correlation_Function_2}.

By substituting $E(r)$ into Eq.\ \eqref{eq:Correlation_Simplified_KFPM}, we derive the correlation function to describe the KPFM-measured potential with a lateral resolution of $\ell_r$:
\begin{equation}
	C(r)=\frac{V_{rms}^2}{(1+\frac{\alpha \ell_r}{\lambda})^2} \times \left( 1 - \frac{\frac{r}{\lambda+\alpha \ell_r} \times \left( 4 - \frac{r}{\lambda + \alpha \ell_r} \right)}{\pi} \right),
	\label{eq:F_z}
\end{equation}
where the original patch potential is characterized by a side length of $\lambda$ and an RMS value of $V_{rms}$.

Using this correlation function and Eq.\ \eqref{eq:Patch_Force_Equation}, we derive the formula for the electrostatic patch force under the influence of KPFM measurements, which can be expressed as
\begin{widetext}
	\begin{equation}
		F_z = - \frac{\epsilon_0 V_{rms}^2}{(1+\frac{\alpha \ell_r}{\lambda})^2 z_{pp}^2} \int_0^{\infty} \frac{k}{\sinh^2 k} \left( \int_0^{k \frac{\lambda}{z_{pp}}(1+\frac{\alpha \ell_r}{\lambda})} r J_0(r) \left( 1-\frac{\frac{r}{k \frac{\lambda}{z_{pp}}(1+\frac{\alpha \ell_r}{\lambda})}\times (4 - \frac{r}{k \frac{\lambda}{z_{pp}}(1+\frac{\alpha \ell_r}{\lambda})})}{\pi}\right) \diff r \right) \diff k.
		\label{eq:Force_z_KPFM}
	\end{equation}
\end{widetext}

This formula offers a straightforward method to estimate the magnitude of patch force when $\ell_r < \lambda$, provided that the patch characteristic size $\lambda$, potential variance $V_{rms}$, plate separation $z_{pp}$ and KPFM lateral resolution $\ell_r$ are known. Notably, the integral depends only on the ratios of $z_{pp} / \lambda$ and $\ell_r / \lambda$, allowing the length parameters to be expressed relative to $\lambda$. It also implies that the behavior of the patch force with respect to $z_{pp}$ and $\ell_r$ can be scaled up or down for different values of $\lambda$.

\begin{figure*}
	\includegraphics{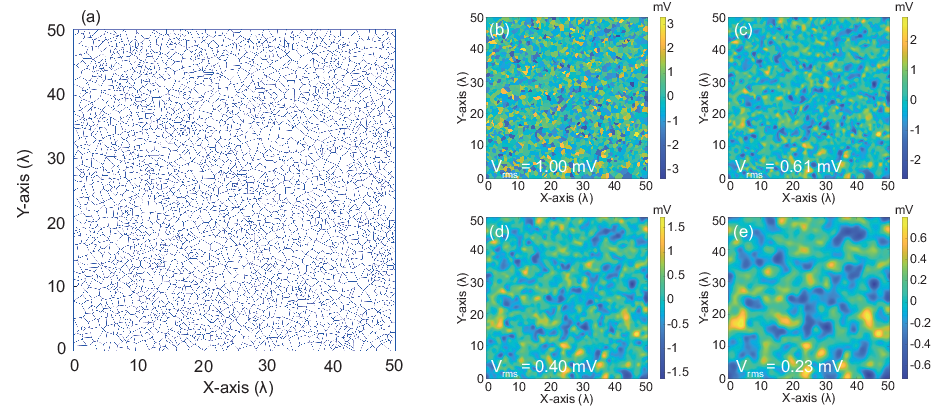}
	\caption{
		Simulation of KPFM measurements.
		(a) The Voronoi diagram illustrating patches with a patch size of $\lambda$ = 50\,nm.
		(b) The original patch potential generated with a patch size of $\lambda$ = 50\,nm and $V_{rms}$ = 1 mV.
		(c, d, e) The corresponding KPFM measurement results with resolutions of 0.28 $\lambda$, 0.57 $\lambda$ and 1.13 $\lambda$, with detailed parameters in Ref.\ \cite{Para}. The RMS value of the patch potential is reduced from 1 mV to 0.23 mV for a resolution of 1.13 $\lambda$.
	}
	\label{fig:Voronoi&KPFM_Potential}
\end{figure*}

\section{Numerical Simulation and Discussions}

In this section, we assess the impact of KPFM measurements on the patch force evaluation using two approaches. First, we use numerical simulations to obtain the KPFM transfer function and calculate the patch force, which {allows us to employ a potential distribution that closely mimics real conditions. Second, we apply the previously derived correlation function to calculate the patch force.

\subsection{Potential Measurement with KPFM}

The potential measured by KPFM $\varphi(\vec{x}_t)$ can be modeled as a convolution of a transfer function $h(\vec{x})$ with the original potential $v(\vec{x})$ \cite{jacobs_resolution_1998}:
\begin{equation}
	\begin{aligned}
			& \varphi \left(x_t,y_t\right) = \iint h(x-x_t,y-y_t) v(x,y) \diff x \diff y , \\
			& h(x-x_t,y-y_t) = \lim_{\Delta x,\Delta y \rightarrow 0} \left( \frac{K^{'}(x-x_t,y-y_t)}{K_{tot}^{'} \Delta x \Delta y} \right),
		\end{aligned}
	\label{eq:KPFM_Potential}
\end{equation}
where $ K'(x-x_t,y-y_t)$ is the derivative of the capacitance between the tip at $(x_t, y_t)$ and an infinitely small surface element $\Delta x \times \Delta y$ at  $(x, y)$, and $K'_{tot}$ is the derivative of the total tip–surface capacitance.

Before evaluating the electrostatic patch forces, let us investigate the effects of convolution through simulations of KPFM measurements. A Voronoi diagram is used to simulate the random patch configuration under the quasi-local correlation model [Fig.\ \ref{fig:Voronoi&KPFM_Potential}(a)]. As an example, we first generate an original potential distribution on a Voronoi diagram with a patch characteristic size $\lambda$ of 50\,nm, where each patch is assigned a random potential with a $V_{rms} $ of 1 mV, as shown in Fig.\  \ref{fig:Voronoi&KPFM_Potential}(b). For a given set of tip's apex radius ($R$) and tip-sample separation ($H$), finite element simulation is employed to calculate the capacitances between the tip and all surface elements. This is achieved by calculating the charge of the tip $Q_t$  when setting the potential of the \textit{i}-th surface element to $v_i$, and then using $K(x_i-x_t,y_i-y_t)$ = $-Q_t/v_i$ \cite{jacobs_resolution_1998}. With the calculated capacitances, we derive the transfer function $h(\vec{x})$ . Subsequently, we perform a two-dimensional convolution of the transfer function with the original potential. 

Fig.\ \ref{fig:Voronoi&KPFM_Potential} (c), (d), (e) show the simulated KPFM measurement results using three different sets of tip parameters. The results show that the KPFM-measured potentials are smeared out compared to the original potentials. Additionally, as $\ell_r$ increases, the measured patch characteristic size expands, while the potential variance decreases.

\subsection{Evaluation of the Electrostatic Patch Force}

\begin{figure*}
	\includegraphics{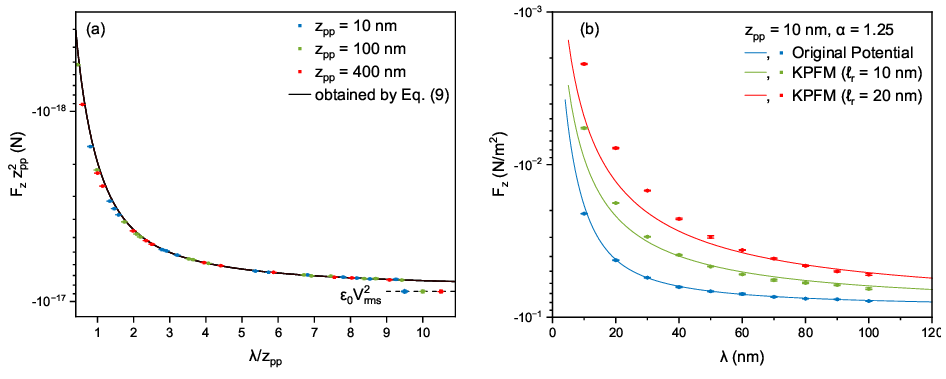}
	\caption{
		The electrostatic patch force calculated using analytic equations (lines) and numerical simulations (dots). Each data point includes four repeated runs with randomly generated patch potentials, with error bars denoting the standard errors. (a) The $F_z z_{pp}^2$ as a function of $\lambda / z_{pp}$, calculated based on the original patch potentials. The RMS value of the potential is set to 1 mV. (b) The electrostatic patch force as a function of $\lambda$, calculated using both original patch potentials and KPFM-measured potentials. The RMS value of the original potential is set to 1 mV.
	}
	\label{fig:Empirical_Validation}
\end{figure*}

We begin by generating a random surface potential distribution following the procedure presented in the previous subsection. With the generated potential  $V(\vec{x})$, we numerically calculate its Fourier coefficient $V(\vec{k})$ and then determine the autocorrelation function of the patch potentials in $k$-space, $C(\vec{k})=V(\vec{k})V(-\vec{k})$ \cite{speake_forces_1996}. We then apply Eq.\  \eqref{eq:Patch_Force_Equation} to calculate the electrostatic patch force in the plate-plate model.

The forces obtained from the numerical simulations are compared with those calculated using the correlation function [Eq.\ \eqref{eq:Force_z}], plotted as $F_z z_{pp}^2$ vs. $\lambda / z_{pp}$ in Fig.\ \ref{fig:Empirical_Validation}(a). The results are consistent with each other across different parameters of $\lambda$ and $z_{pp}$, indicating that the correlation function captures the main features of the patch potential distribution, although the effects of random patch size and shape are ignored in its derivation. Additionally, we observe that the magnitude of the force rapidly diminishes when $\lambda$ is smaller compared to $z_{pp}$ while reaches the limit of $\epsilon_0 V_{rms}^2 / z_{pp}^2$ when $\lambda$ is much greater than $z_{pp}$.

To account for the effects of KPFM measurements, the KPFM-measured potentials are used to calculate $V(\vec{k})$ and subsequently the electrostatic patch forces. These KPFM-measured potentials are obtained following the numerical method described in Section III A. The results, shown in Fig.\  \ref{fig:Empirical_Validation}(b), are compared with the results calculated using the correlation function for KPFM-measured potentials [Eq.\ \eqref{eq:Force_z_KPFM}]. Due to the invalidity of the boundary zone assumption, the theoretical lines deviate from simulation data points when $\lambda < \ell_r$. Generally, the electrostatic patch forces exhibit a similar dependence on $\lambda$ for both the original patch potential and the KPFM-measured potential. The patch forces calculated from the KPFM-measured potential are consistently smaller than those from the original potential. As $\lambda$ becomes sufficiently large, the forces evaluated with different resolutions converge to the same limit, $\epsilon_0 V_{\text{rms}}^2 / z_{pp}^2$.

To quantify the effects of KPFM measurements on estimating electrostatic patch forces, we define $\eta$ as the ratio of the force magnitude derived from the KPFM measurements to that calculated directly from the original generated potentials. We assess this metric as a function of the lateral resolution $\ell_r$ under different plate-plate separations $z_{pp}$, as shown in Fig.\ \ref{fig:Eta_Display}. Both $\ell_r$ and $z_{pp}$ are in unit of $\lambda$. We use $\lambda$ = 80 nm as an example, and similar results are obtained with larger $\lambda$. 

\begin{figure}
	\includegraphics{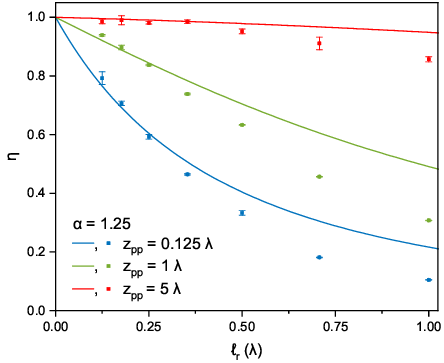}   
	\caption{
		Values of $\eta$ as a function of $\ell_r$ for different $z_{pp}$. Dots represent numerical simulations, and lines represent calculations based on analytic formulas. Each data point includes four repeated runs with randomly generated patch potentials, with error bars denoting the standard errors. The RMS value of the original potential is 1 mV. The data shown here use $\lambda = 80$ nm, and similar results are obtained for $\lambda = 0.8$ mm. 
	}
	\label{fig:Eta_Display}
\end{figure}

It is found that as $\ell_r$ increases, the electrostatic patch force is increasingly underestimated. The behavior of the underestimation depends on the ratio of $z_{pp}/\lambda$. When $z_{pp} \gg \lambda$, $\eta$ becomes less dependent on $\ell_r$, allowing for accurate patch force evaluation even with relatively low lateral resolution, which is relevant for the situation of patch force evaluation in space-based gravitational wave detection. The reason is that when the separation between the surfaces is large relative to the patch size, the patch force is significantly diminished by an averaging effect—each patch on one surface interacts with multiple patches on the opposing surface, effectively averaging out the force. In this case, the patch force calculation is less sensitive to the KPFM averaging effects.

To understand the underestimation of patch force evaluation when the lateral resolution exceeds the patch characteristic size, where the formula we derived is invalid, we perform a rough analysis using the data reported in Ref. \cite{robertson_kelvin_2006} as a reference. In that work, a TiC/Ti sample was scanned at a distance of 100 $\mu$m using a cylindrical probe tip with a diameter of 3 mm, corresponding to a lateral resolution of $\ell_r \sim 1.26$ mm, derived using the method described in Appendix \ref{appendix:linearity}. Based on Fig. 3 in Ref. \cite{robertson_kelvin_2006}, we estimate the measured patch size to be  $\lambda \sim 1.4$ mm and the potential variance to be $V_{rms} \sim 1.28$ mV. Given a test mass-housing separation of $z_{pp} = 2.9$ mm, the patch force is estimated to be $-1.04 \times 10^{-13}$ N/m$^2$ using Eq.\ \eqref{eq:Force_z}, if we assume that the measured values represent the actual potential distribution. However, due to the averaging effect of KPFM, the actual patch size is likely to be much smaller, with a correspondingly larger potential variance. Using the method described in Sec. III A, we numerically generate two sets of potential distributions with the following statistical parameters: $V_{rms} = 14.7$ mV, $\lambda$ = 0.4 mm; and $V_{rms} = 7.6$ mV, $\lambda$ = 0.8 mm.  After simulating the KPFM measurement results for both parameter sets, the resulting statistical parameters match those measured in Ref. \cite{robertson_kelvin_2006}. Based on these parameters, we calculate the patch force to be $-1.26 \times 10^{-12}$ N/m$^2$ and $-1.20 \times 10^{-12}$ N/m$^2$, respectively. These forces are approximately an order of magnitude greater than the results estimated using the KPFM-measured potential, but show a slight dependence on $\lambda$.

\section{Conclusion}
KPFM plays a crucial role in characterizing surface potentials, which is essential for accurately estimating electrostatic patch forces in various applications, such as gravitational wave detection and Casimir force experiments. Different types of experiments involve sample configurations of different scales, making it important to determine the appropriate lateral resolution of the KPFM to ensure accurate force evaluation. To address this, we derived two correlation functions for patch potentials under the modified quasi-local correlation model—one for the original patch potential and the other for its KPFM-measured counterpart. Using these correlation functions, we derived formulas to calculate the electrostatic patch forces between two infinite parallel plates, which enable one to quickly assess the validity of the KPFM lateral resolution using $V_{rms}$, $\lambda$ and $\ell_r$ as inputs. We further conducted numerical simulations to investigate the dependence of patch force evaluation on the lateral resolution of KPFM with a more realistic patch potential distribution. 

Results from both analytic method and numerical simulation show that the accuracy of force evaluation depends on the ratios of the sample separation and the KPFM's resolution relative to the patch characteristic size. Due to the averaging effect inherent in KPFM measurements, the estimated patch forces tend to be smaller than their actual magnitude, but become less sensitive to the lateral resolution when the plate-plate separation is much larger than the patch size. These findings could provide valuable guidance on determining the necessary lateral resolution of KPFM in different experiments. Future work could explore force estimation methods when the lateral resolution is larger than the patch size, as well as force estimation using different potential distribution models. It is also valuable to experimentally test the dependence on the lateral resolution of KPFM by performing measurements on control samples using different probes with various lateral resolutions. Furthermore, developing methods to accurately recover the true potential from KPFM measurements remains an important avenue for future research.

\begin{acknowledgments}
This work was supported by the National Key R\&D Program of China (Grant No. 2022YFC2204100) and the National Natural Science Foundation of China (NSFC) (Grant No. 11875137).
\end{acknowledgments}

\appendix

\section{Derivation of the Correlation Function}
\label{appendix:Derivation}

With the surface potential described by Eq.\ \eqref{eq:Surface_Potentail_KPFM} and $\left< v_a v_b \right>_v = \delta_{ab} V_{rms}^2$, the correlation function can be derived as follows:
\begin{equation*}
	\begin{aligned}
		C(\vec{x},\vec{x}^{'}) &= \left< V(\vec{x}) V(\vec{x}^{'}) \right>_v \\
		&= \frac{\sum_{a,b} \left< v_a v_b \right>_v \Theta_a (\vec{x}) \Theta_b (\vec{x}^{'})}{\left( \sum_c \Theta_c(\vec{x}) \right) \cdot \left( \sum_d \Theta_d(\vec{x}^{'}) \right)} \\
		&= V_{rms}^2 \cdot \frac{\sum_a \Theta_a(\vec{x}) \Theta_a(\vec{x}^{'})}{\left( \sum_c \Theta_c(\vec{x}) \right) \cdot \left( \sum_d \Theta_d(\vec{x}^{'}) \right)} \\
		&= V_{rms}^2 \cdot \frac{\sum_a \Theta_a(\vec{x}) \Theta_a(\vec{x}^{'})}{\left( \sum_c \Theta_c(\vec{x}) \Theta_c(\vec{x}) \right) \cdot \left( \sum_d \Theta_d(\vec{x}^{'}) \Theta_d(\vec{x}^{'}) \right)}
	\end{aligned}
\end{equation*}
where $\sum_a \Theta_a (\vec{x}) = \sum_a \Theta_a(\vec{x}) \Theta_a (\vec{x})$ is used in the last step.  According to the definition of the patch pair coverage function, the correlation function is rewritten as
\begin{equation*}
	C(\vec{x},\vec{x}^{'}) = V_{rms}^2 \cdot \frac{E(\vec{x}, \vec{x}^{'})}{E(\vec{x}, \vec{x})E(\vec{x}^{'}, \vec{x}^{'})}.
\end{equation*}

Assuming translation and rotation invariance, $E(\vec{x}, \vec{x}^{'})$ is reduced to  $E(r)$ with $\vec{r} = \vec{x} - \vec{x}^{'}$. Then the correlation function can be expressed as
\begin{equation*}
	C(r) = V_{rms}^2 \cdot \frac{E(r)}{E^2(0)}.
\end{equation*}

\renewcommand{\thefigure}{B1}
\begin{figure}
	\includegraphics{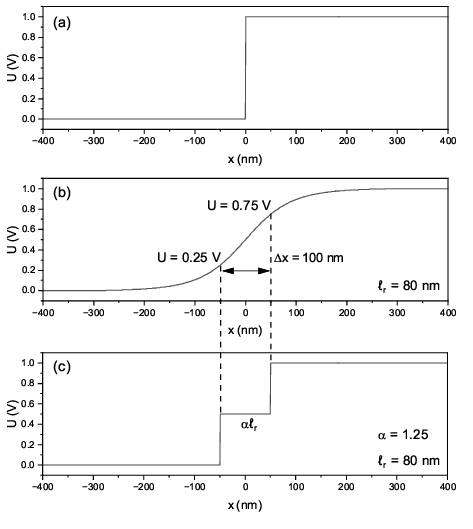}
	\caption{
		One-dimensional potential distribution: (a) Ideal step potential; (b) KPFM-measured potential of the step potential shown in (a); (c) Simplified model of the KPFM-measured potential with a boundary zone of a single averaged value.
	}
	\label{fig:alpha}
\end{figure}

\renewcommand{\thefigure}{B2}
\begin{figure}
	\includegraphics{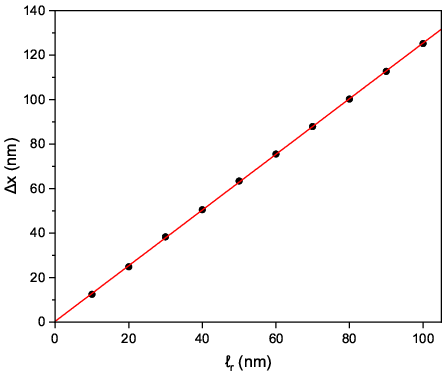}   
	\caption{
		Relation between $\Delta x$ and the KPFM's resolution $\ell_r$. Solid dots represent the simulation data, the solid line is a linear fit with a slope of 1.25 and an intercept of 0.36 nm.
	}
	\label{fig:Linearity}
\end{figure}

\section{Definition of $\alpha$}
\label{appendix:linearity}

The coefficient $\alpha$ is introduced to account for the difference between the definitions of the KPFM's lateral resolution and the width of the boundary zone. In Ref.\ \cite{leveque_sensitivity_2005}, the lateral resolution is defined as the distance over which the phase signal of KPFM changes from $\frac 1 4$ to $\frac 3 4$ of the maximum variation. It has been shown that when the separation $H$ is much smaller than the tip's apex radius $R$, the lateral resolution $\ell_r$ can be approximated by $\ell_r \approx \sqrt{R H}$ \cite{leveque_sensitivity_2005}. In our numerical simulations, all KPFM parameters satisfy this condition, so we directly use this approximation as the lateral resolution.

In our work, we apply a similar definition to determine the boundary zone width. A one-dimensional step potential, as shown in Fig.\ \ref{fig:alpha}(a), is introduced for this purpose. The convolution of this step potential with the transfer function produces the KPFM-measured potential, shown in Fig.\ \ref{fig:alpha}(b). In this measured potential, the value transitions smoothly from $U_1$ to $U_2$. We define the intermediate interval, $\Delta x$, as the distance over which the measured potential shifts from $\frac{1}{4}$ to $\frac{3}{4}$ of the total potential variation, as illustrated in Fig.\ \ref{fig:alpha}(b). The boundary zone width, $\alpha \ell_r$, is then set to match this intermediate interval, as shown in Fig.\ \ref{fig:alpha}(c). Our analysis reveals that $\alpha \approx 1.25$, and this relationship holds consistently when the resolution varies. It should be noted that the value of $\alpha$ depends on the choice of upper and lower ratios for defining the interval.

To verify the consistency of $\alpha$ across various KPFM resolutions, we calculated the values of $\Delta x$ for different $\ell_r$ following the same procedure, as shown in Fig.\ \ref{fig:Linearity}. The linear relationship between $\Delta x$ and $\ell_r$ is evident, with the slope estimated to be $\alpha \approx 1.25$. This value of $\alpha$ was used in all simulations.

%

\end{document}